\documentclass[aps,frontiers,twocolumn,showpacs,superscriptaddress,groupedaddress]{revtex4-1}

\usepackage{url,lineno}
\usepackage{mathrsfs}
\usepackage{units}
\usepackage{bm}
\usepackage[usenames,dvipsnames]{color}
\usepackage{dcolumn}
\usepackage[usenames,dvipsnames,svgnames,table]{xcolor}
\usepackage{amssymb}
\usepackage{amsmath}
\usepackage{amsfonts}
\usepackage{graphicx}
\usepackage{xstring}
\usepackage{units}

\renewcommand{\v}[1]{\mathbf{#1}}
\newcommand{\expect}[1]{\left< #1 \right>}

\newcommand{\Cm}{C_\mathrm{m}}
\newcommand{\gl}{g_\mathrm{l}}
\newcommand{\El}{E_\mathrm{l}}
\newcommand{\Isyn}{I^\mathrm{syn}}
\newcommand{\Iext}{I^\mathrm{ext}}
\newcommand{\Erev}{E^\mathrm{rev}}

\newcommand{\wsyn}{w^\mathrm{syn}}
\newcommand{\tausyn}{\tau^\mathrm{syn}}
\newcommand{\nusyn}{\nu^\mathrm{syn}}
\newcommand{\gtot}{g^\mathrm{tot}}
\newcommand{\tauref}{\tau^\mathrm{ref}}
\newcommand{\Mexc}{M_\mathrm{exc}}
\newcommand{\Minh}{M_\mathrm{inh}}
\newcommand{\ueff}{u_\mathrm{eff}}
\newcommand{\taueff}{\tau_\mathrm{eff}}
\newcommand{\DKL}{D^\mathrm{norm}_\mathrm{KL}}
\newcommand{\pa}{\mathrm{\textbf{pa}}}
 
\DeclareMathOperator\unif{unif}

\newenvironment{newstuff}
{\color{black}}
{\color{black}}

\begin{document}

\title{Probabilistic inference in discrete spaces can be implemented into networks of LIF neurons}
\thanks{
The first two authors contributed equally to this work.
We thank W.~Maass for his essential support.
This research was supported by EU grants \#269921 (BrainScaleS), \#604102 (Human Brain Project), the Austrian Science Fund FWF \#I753-N23 (PNEUMA) and the Manfred St\"ark Foundation.
We acknowledge financial support of the Deutsche Forschungsgemeinschaft and Ruprecht-Karls-Universität Heidelberg within the funding program Open Access Publishing.
}

\author{Dimitri Probst\,$^{1,*}$, Mihai A.\ Petrovici\,$^{1,*}$, Ilja Bytschok\,$^{1}$,
    Johannes Bill\,$^{2}$, Dejan Pecevski\,$^{2}$, Johannes Schemmel\,$^{1}$ and Karlheinz Meier\,$^{1}$}
\affiliation{$^1$Kirchhoff Institute for Physics, University of Heidelberg, Heidelberg, Germany \\ $^2$Institute for Theoretical Computer Science, Graz University of Technology, Graz, Austria}

\date{\today}

\begin{abstract}

The means by which cortical neural networks are able to efficiently solve inference problems remains an open question in computational neuroscience. Recently, abstract models of Bayesian computation in neural circuits have been proposed, but they lack a mechanistic interpretation at the single-cell level. In this article, we describe a complete theoretical framework for building networks of leaky integrate-and-fire neurons that can sample from arbitrary probability distributions over binary random variables. We test our framework for a model inference task based on a psychophysical phenomenon (the Knill-Kersten optical illusion) and further assess its performance when applied to randomly generated distributions. As the local computations performed by the network strongly depend on the interaction between neurons, we compare several types of couplings mediated by either single synapses or interneuron chains. Due to its robustness to substrate imperfections such as parameter noise and background noise correlations, our model is particularly interesting for implementation on novel, neuro-inspired computing architectures, which can thereby serve as a fast, low-power substrate for solving real-world inference problems.

\end{abstract}

\maketitle

\section{Introduction}
\label{s:int}

The ability of the brain to generate predictive models of the environment based on sometimes ambiguous, often noisy and always incomplete sensory stimulus represents a hallmark of Bayesian computation.
Both experimental \citep{yang2007probabilistic,berkes11} and theoretical studies \citep{rao2004hierarchical,deneve2008bayesian,buesing2011neural} have explored this highly intriguing but also hotly debated hypothesis.
These approaches have, however, remained rather abstract, employing highly idealized neuron and synapse models.

In this study, we explore how recurrent networks of leaky integrate-and-fire (LIF) neurons -- a standard neuron model in computational neuroscience -- can calculate the posterior distribution of arbitrary Bayesian networks over binary random variables through their spike response.
Our work builds upon the findings of three previous studies:
In \citet{buesing2011neural}, it was shown how the spike pattern of networks of abstract model neurons can
be understood as Markov Chain Monte Carlo (MCMC) sampling from a well-definded class of target distributions.
The approach was extended in a follow up paper \citep{pecevski11} to Bayesian networks by identifying appropriate network architectures.
The theoretical foundation for taking the step from abstract neurons to more realistic networks of LIF neurons was developed recently in \citet{petrovici13lifsampling}.
In this paper, we follow and extend the approach from \citet{petrovici13lifsampling} to the network architectures proposed by \citet{pecevski11}.
In particular, we describe a blueprint for designing spiking networks that can perform sample-based inference in arbitrary graphical models.

We thereby provide the first fully functional implementation of Bayesian networks with realistic neuron models.
This enables studies in two complementary fields.
On one hand, the development of network implementations for Bayesian inference contributes to the open debate on its biological correlate by exploring possible realizations in the brain.
These can subsequently guide both targeted experimental research and computational modeling.
Furthermore, additional physiological investigation is now made possible, e.g., of the influence of specific neuron and synapse parameters and dynamics or the embedding in surrounding networks.
On the other hand, the finding that networks of LIF neurons can implement parallelized inference algorithms provides an intriguing application field for novel computing architectures.
Much effort is currently invested into the development of neuro-inspired, massively parallel computing platforms, called neuromorphic devices \citep{indiveri_tnn2006,schemmel_iscas2010,furber2013}.
These devices typically implement models of LIF neurons which evolve in parallel and without a central clock signal.
This paper offers a concrete concept for the application of neuromorphic hardware as powerful inference machines.
Interestingly, questions similar to the ones mentioned above in a biological context arise for artificial systems as well: the effect of parameter noise or limited bandwidth on functional network models is, for example, the subject of active research \citep{petrovici2014characterization}.

The document is structured as follows.
In Sec.~\ref{sec:theory}, we review and adapt the theories from \citet{buesing2011neural,pecevski11} and \citet{petrovici13lifsampling} to build a complete framework for embedding sampling from probability distributions into the structure and dynamics of networks of LIF neurons.
In Sec.~\ref{s:res}, we provide the required translation rules and demonstrate the feasibility of the approach in computer simulations.
We further compare the effect of different synaptic coupling dynamics and present a mechanism based on interneuron chains which significantly improves the sampling quality. 
Finally, we study robustness to parameter distortions and to correlations in the background noise, 
as these are likely to be present in any physical substrate, be it biological or artificial.
\begin{newstuff}
    In Sec.~\ref{s:con}, we discuss these results and their implications for biological and neuro-inspired computing architectures.
\end{newstuff}

For the simulations with LIF neurons, we used PyNN \citep{davison08pynn} with NEST \citep{diesmann01nest} or NEURON \citep{hines03neuron} as back-end.
The simulations with networks of abstract model neurons were conducted in Python.

\section{Materials \& Methods}
\label{sec:theory}

\subsection{Bayesian Networks as Boltzmann Machines}
\label{ss:bn}
        
The joint distribution defined by a Bayesian graph is the product of conditional distributions, one for each random variable (RV), with its value conditioned on the values of its parent variables.
For a graph with $K$ binary RVs $Z_k$, the joint probability distribution is given by
\begin{equation}
    \label{graph}
    p(\mathbf{Z} = \mathbf{z}) =: p(\mathbf{z}) = \prod_{k = 1}^K \frac{1}{Z} \Phi_k(\mathbf{z}_k) := \prod_{k = 1}^K p(z_k | \mathrm{\textbf{pa}}_k) \quad ,
\end{equation}
where $\mathbf{z}_k$ represents the state vector of the variables $\mathbf{Z}_k$ in $\Phi_k$, which we henceforth call principal RVs, and $\mathrm{\textbf{pa}}_k$ represents the state vector of the parents of $Z_k$.
$Z$ is a normalizing constant; w.l.o.g., we assume $\Phi_k>1$.
The factor $p(z_k | \mathrm{\textbf{pa}}_k)$ is called an $n^\mathrm{th}$-order factor if it depends on $n$ RVs or rather $|\mathrm{pa}_k| = n - 1$.

\begin{figure*}[tbp]
    \centering
    \includegraphics[width=0.95\linewidth]{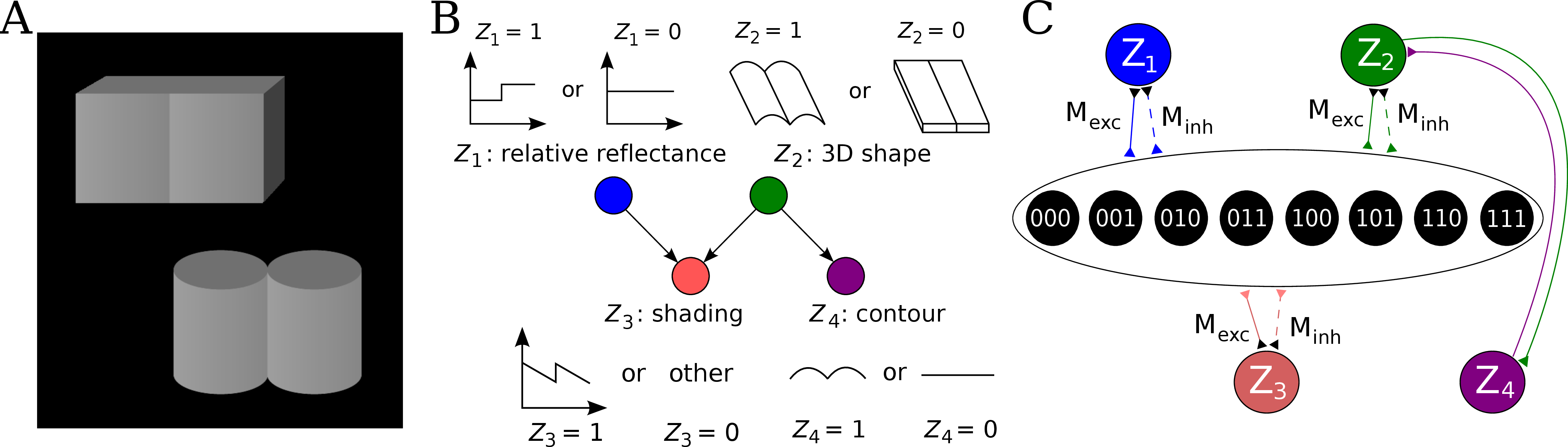}
    \caption{
         Formulation of an example inference problem as a Bayesian network and translation to a Boltzmann machine.
        (\textbf{A}) Knill-Kersten illusion from \cite{knill91}.
            Although the four objects are identically shaded, the left cube is perceived as being darker than the right one.
            This illusion depends on the perceived shape of the objects and does not occur for, e.g., cylinders.
        (\textbf{B}) The setup can be translated to a Bayesian network with four binary RVs.
            The (latent) variables $Z_1$ and $Z_2$ encode the (unknown) reflectance profile and 3D shape of the objects, respectively.
            Conditioned on these variables, the (observed) shading and 2D contour are encoded by $Z_3$ and $Z_4$, respectively.
            \begin{newstuff}
                Figure modified from \citet{pecevski11}.
            \end{newstuff}
        (\textbf{C}) Representation of the Bayesian network from \textbf{B} as a Boltzmann machine.
            Factors of order higher than 2 are replaced by auxiliary variables as described in the main text.
            The individual connections with weights $\Mexc,\; \Minh \to \infty$ between each principal and auxiliary variable have been omitted for clarity.
        \label{fig:kki}
    }
\end{figure*}

Such a Bayesian network can be transformed into a second-order Markov random field (i.e., an MRF with a maximum clique size of 2).
Here, we follow the recipe described in \citet{pecevski11}.
First and second-order factors are easily replaceable by potential functions $\Psi_k(Z_k)$ and $\Psi_k(Z_{k1}, Z_{k2})$, respectively.
For each $n^\mathrm{th}$-order factor $\Phi_k$ with $n>2$ principal RVs, we introduce $2^n$ auxiliary binary RVs $X_k^{\mathbf{z}_k\in\mathscr{Z}_k}$, where $\mathscr{Z}_k$ is the set of all possible assignments of the binary vector $\mathbf{Z}_k$ (Fig. \ref{fig:kki} C).
Each of these RVs ``encode'' the probability of a possible state $\mathbf{z}_k$ within the factor $\Phi_k$ by introducing the first-order potential functions $\Psi_k^{\mathbf{z}_k}(X_k^{\mathbf{z}_k} = 1) = \Phi_k(\mathbf{Z}_k = \mathbf{z}_k)$.
The factor $\Phi_k(\mathbf{Z}_k)$ is then replaced by a product over potential functions
\begin{equation}
    \Phi_k(\mathbf{Z}_k) = \prod_{\mathbf{z}_k} \Psi_k^{\mathbf{z}_k}(X_k^{\mathbf{z}_k}) \prod_{i=1}^n \chi_{ki}^{\mathbf{z}_k}(Z_{ki}, X_k^{\mathbf{z}_k}) \quad ,
\end{equation}
where an auxiliary RV $X_k^{\mathbf{z}_k}$ is active if and only if the principal RVs $\mathbf{Z}_k$ are active in the configuration $\mathbf{z}_k$.
Formally, this corresponds to the assignment: $\chi_{ki}^{\mathbf{z}_k}(Z_{ki}, X_k^{\mathbf{z}_k}) = 1 - X_k^{\mathbf{z}_k} (1 - \delta_{Z_{ki},z_{ki}})$.
In the graphical representation, this amounts to removing all directed edges within the factors and replacing them by undirected edges from the principal to the auxiliary RVs.
It can then be verified \citep{pecevski11} that the target probability distribution can be represented as a marginal over the auxiliary variables.

As the resulting graph is a second-order MRF, its underlying distribution can be cast in Boltzmann form:
\begin{equation}
    p(\mathbf{z,x}) = \frac{1}{Z}\exp \left( \frac{1}{2} \mathbf{z}^T \mathbf{W} \mathbf{z} + \frac{1}{2} \mathbf{z}^T \mathbf{V} \mathbf{x} + \mathbf{z}^T \mathbf{b} + \mathbf{x}^T \mathbf{a} \right) \quad ,
    \label{eq:boltzmann}
\end{equation}
where the (symmetric) weight matrices $\mathbf{W, V}$ and bias vectors $\mathbf{b,a}$ are defined as follows:
\begin{align}
    W_{Z_{ki},Z_{kj}} \quad &= \quad \left\{\begin{array}{ll}
        \log\frac{\Phi_k\left(Z_{ki}=0,Z_{kj}=0\right)\Phi_k\left(Z_{ki}=1,Z_{kj}=1\right)}{\Phi_k\left(Z_{ki}=0,Z_{kj}=1\right)\Phi_k\left(Z_{ki}=1,Z_{kj}=0\right)} \quad \\ \quad \text{within second-order factors} \; \Phi_k \\
        0 \quad \text{otherwise} \end{array}\right. \\
    V_{Z_{ki},X_k^{\mathbf{z}_k}} \quad &= \quad \left\{\begin{array}{ll}
        \Mexc \quad \text{if} \; \mathbf{z}_{ki} = 1 \\
        \Minh \quad \text{if} \; \mathbf{z}_{ki} = 0 \end{array}\right.
        \label{weight_matrix_v}
\end{align}
\begin{align}
    b_{Z_{ki}} \quad &= \quad \left\{\begin{array}{ll} 
        \log\frac{\Phi_k\left(Z_{ki}=1\right)}{\Phi_k\left(Z_{ki}=0\right)} \quad \text{within first-order factors} \\
        \log\frac{\Phi_k\left(Z_{ki}=1,Z_{kj}=0\right)}{\Phi_k\left(Z_{ki}=0,Z_{kj}=0\right)} \quad \\ \quad \text{within second-order factors}\end{array}\right.\\
    a_{X_k^{\mathbf{z}_k}} \quad &= \quad  \log \left(\Phi_k - 1\right) - L^1(\mathbf{z}_k) \Mexc \quad ,
\end{align}
all other matrix and vector elements being zero.
$L^1(\cdot)$ represents the $L^1$ norm.
In the theoretical model, $\Mexc=\infty$ and $\Minh=-\infty$, but they receive finite values in the concrete implementation (Sec. \ref{ss:aux}).
From here, it is straightforward to create a corresponding classical Boltzmann machine.
We therefore use a simplified notation from here on: we consider the vector $\v{Z}$ to include both principal and auxiliary RVs and the Boltzmann distributions over $\v{Z}$ are henceforth defined by the block diagonal weight matrix $\v{W}$ and the bias vector $\v{b}$.

\subsection{Neural Sampling: An Abstract Model}
\label{sec:neuralsampling}

Gibbs sampling is typically used to update the states of the units in a Boltzmann machine.
However, in a spiking network, detailed balance is not satisfied, since spiking neurons do not incorporate reversible dynamics due to the existence of refractory mechanisms.
While a non-refractory neuron can always be brought into the refractory state with sufficient stimulation, the reverse transition is, in general, not possible.
It is possible, however, to understand the dynamics of a network of stochastic neurons as MCMC sampling.
In the following, we use the model proposed in \citet{buesing2011neural} for sampling from Boltzmann distributions (Eq.~\ref{eq:boltzmann}).

\begin{figure*}[tbp]
    \centering
    \includegraphics[width=1.0\linewidth]{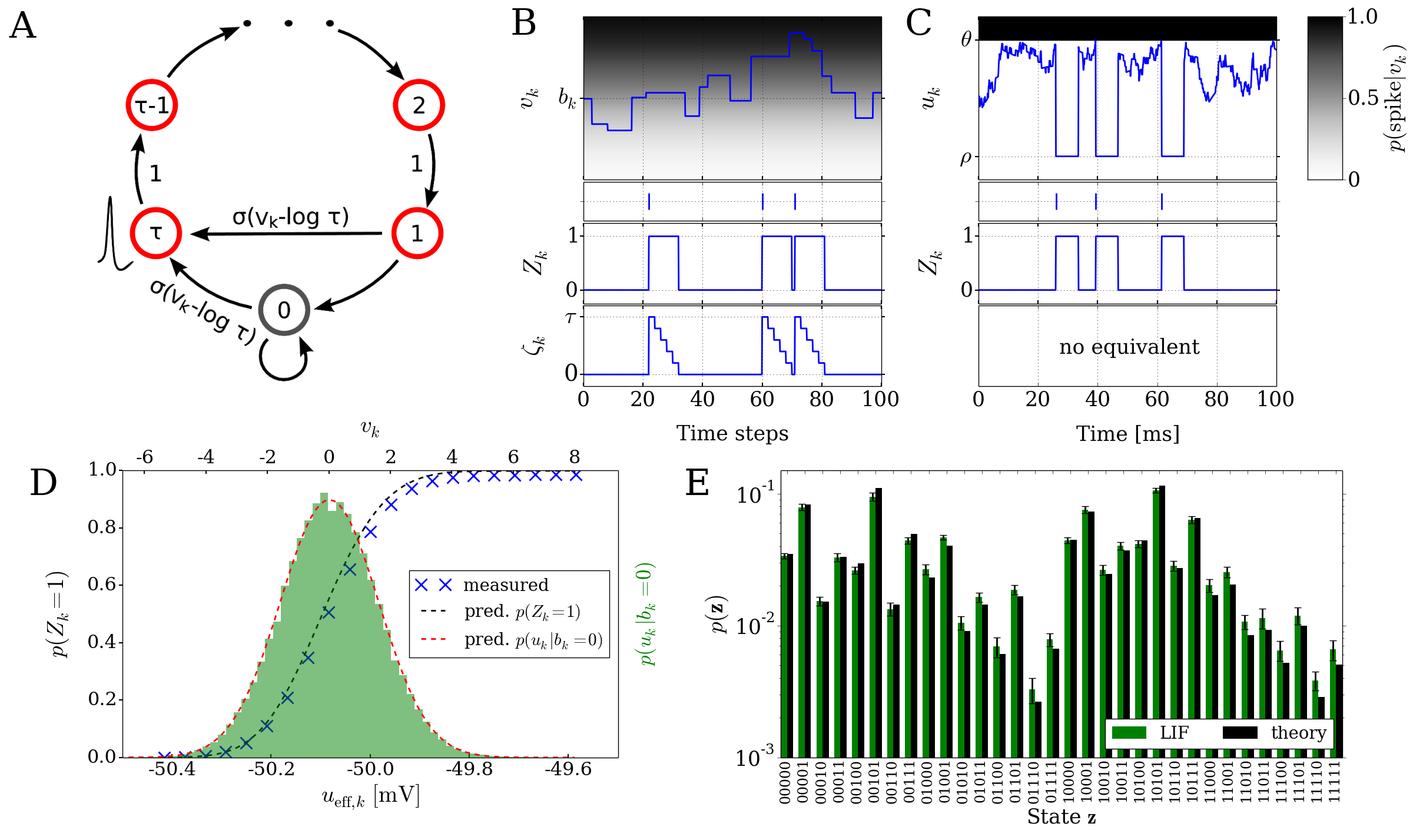}
    \caption{
         Neural sampling: abstract model vs. implementation with LIF neurons.
        (\textbf{A}) Illustration of the Markov chain over the refractory variable $\zeta_k$ in the abstract model.
        \begin{newstuff}
            Figure taken from \citet{buesing2011neural}.
        \end{newstuff}
        (\textbf{B}) Example dynamics of all the variables associated with an abstract model neuron.
        (\textbf{C}) Example dynamics of the equivalent variables associated with an LIF neuron.
        (\textbf{D}) Free membrane potential distribution and activation function of an LIF neuron: theoretical prediction vs.~experimental results.
        The {\it blue} crosses are the mean values of $5$ simulations of duration $\unit[200]{s}$.
        The error bars are smaller than the size of the symbols.
        Table \ref{tab:cobalif} lists the used parameter values of the LIF neuron.
        (\textbf{E}) Performance of sampling with LIF neurons from a randomly chosen Boltzmann distribution over 5 binary RVs.
        Both weights and biases are chosen from a normal distribution
        $\mathcal{N}\left(\mu=\unit[0],\, \sigma={0.5}\right)$. The {\it green} bars
        are the results of $10$ simulations of duration $\unit[100]{s}$.
        The error bars denote the standard error.
        \label{fig:neuralsampling}   
    }
\end{figure*}

In this model, the spike response of a neuron is associated to the state $z_k$ of an RV $Z_k$ and a spike is interpreted as a state switch from 0 to 1.
Each spike is followed by a refractory period of duration $\tau$, during which the neuron remains in the state $Z_k=1$.
The so-called neural computability condition (NCC) provides a sufficient condition for correct sampling, wherein a neuron's "knowledge" about the state of the rest of the network - and therefore its probability to spike - is encoded in its membrane potential:
\begin{equation}
        \label{NCC}
        v_k(t) = \log\frac{p(Z_k(t)=1|\mathbf{Z}_{\backslash k}(t))}{p(Z_k(t)=0|\mathbf{Z}_{\backslash k}(t))} \quad ,
\end{equation}
where $\mathbf{Z}_{\backslash k}(t)$ denotes the vector of all other variables $Z_i$ with $i \neq k$.
By solving for $Z_k(t)=1$, one obtains a logistic neural activation function
(Fig.~\ref{fig:neuralsampling} D), which is reminiscent of the update rules in Gibbs sampling:
\begin{equation}
        \label{activation_function}
        p(Z_k(t)=1|\mathbf{z}_{\backslash k}(t)) = \sigma\left(v_k\left(t\right)\right) := \frac{1}{1 + \exp\left(-v_k(t)\right)} \quad ,
\end{equation}
In order for a neuron to be able to track its progression through the refractory period, a refractory variable $\zeta_k$ is introduced for each neuron, which assumes the value $\tau$ following a spike at time $t$ and decreases linearly towards 0 at time $t+\tau$.
The transition probability to the state $\zeta_k$ only depends on the previous state $\zeta_k'$.
The resulting sequence of states $\zeta_k(t=0),\zeta_k(t=1),\zeta_k(t=2),...$ is a Markov chain.
Fig.~\ref{fig:neuralsampling}~A illustrates the transition of the state variable $\zeta_k$.
A neuron with $\zeta_k \in \{0, 1\}$ elicits a spike with probability $\sigma(v_k -\log \tau)$.
This defines the stochastic neuron model in \citet{buesing2011neural}.

For the particular case of a Boltzmann distribution with weight matrix $\mathbf{W}$ and bias vector $\mathbf{b}$, the NCC (Eq. \ref{NCC}) is satisfied by neurons with the membrane potential represented by a sum of rectangular postsynaptic potentials (PSPs):
\begin{equation}
    \label{NCC_bm}
    v_k(t) = b_k + \sum_{i=1}^K W_{ki} Z_i(t) \quad .
\end{equation}
Fig.~\ref{fig:neuralsampling}~B illustrates the time courses of the membrane potential $v_k$, the state variable $Z_k$ and the refractory variable $\zeta_k$ of an abstract model neuron.

\subsection{Neural Sampling with LIF Neurons}
\label{ss:lif}

In contrast to the abstract model described above, biological neurons exhibit markedly different dynamics.
Most importantly, the firing times of individual neurons are not stochastic: in-vitro single neuron experiments show how a fixed stimulus sequence triggers a fixed spike train reliably over multiple trials \citep{mainen1995reliability}.
Also, their interaction is not mediated by rectangular PSPs.
In-vivo, however, neurons often receive diffuse synaptic stimulus that alters their dynamics in several important ways \citep{destexhe03hcs}.
As demonstrated in \citet{petrovici13lifsampling} and described below, under such conditions, deterministic neurons can attain the required stochastic dynamics to sample from arbitrary Boltzmann distributions.

A widely used neuron model that captures the abovementioned characteristics of biological neurons is the leaky integrate-and-fire (LIF) model, which defines neuron membrane dynamics as follows:
\begin{equation}
   \label{eq:lif_cond}
   \Cm \dot{u} (t) = \gl \left[\El - u(t)\right] + \Isyn(t) + \Iext(t) \quad ,
\end{equation}
where $\Cm$, $\gl$ and $\El$ represent the membrane capacitance, the membrane leakage conductance and the membrane leakage potential, respectively, and $\Iext$ represents an external stimulus current.
Whenever $u$ crosses a threshold $\theta$, it is pulled down to a reset value $\rho$, where it stays for refractory time $\tauref$.
For a given $\tauref$ of the LIF neuron and a number $\tau$ of refractory time steps of the abstract model from Sec.~\ref{sec:neuralsampling}, the state interpretation between the two domains can be aligned by interpreting an MCMC update step as a time interval $\Delta t$, such that $\tauref = \tau \; \Delta t$.
The synaptic interaction current $\Isyn$ denotes:
\begin{equation}
   \Isyn(t) = \sum_{\mathrm{syn}\;i} \sum_{\mathrm{spikes}\;s} \wsyn_i \left[ \Erev_i - u(t) \right] \exp \left( \frac{t-t_s}{\tausyn} \right) \quad ,
\end{equation}
where $\wsyn_i$ represents the weight of the $i^\mathrm{th}$ afferent synapse, $\Erev_i$ its reversal potential and $\tausyn$ the synaptic time constant.
Figure \ref{fig:neuralsampling}~C illustrates exemplary time courses of the membrane potential $u_k$ and of the corresponding state variable $Z_k$ of a LIF neuron.

In the regime of diffuse synaptic background noise, it can be shown that the temporal evolution of the membrane potential is well approximated by an Ornstein-Uhlenbeck (OU) process with a mean and variance that can be computed analytically \citep{petrovici13lifsampling}.
This regime is achieved by intense synaptic bombardment from independent Poissonian spike sources with high firing rates $\nusyn$ and low synaptic weights.
This allows calculating the temporal evolution of the membrane potential distribution $p(u|u_0)$, as well as the mean first-passage time $T(u_1,u_2)$ of the membrane potential from $u_1$ to $u_2$.
With this, the activation function of a single LIF neuron can be expressed as
\begin{equation}
    \label{lif_activation_function}
    p(Z_k=1) = \frac{\sum_n n P_n \tauref}{\sum_n P_n (n \tauref + T_n)} \quad ,
\end{equation}
where $P_n$ represents the probability of an $n$-spike burst and $T_n$ the average period of silence following such a burst.
Both of these terms be expressed with recursive integrals that can be evaluated numerically \citep{petrovici13lifsampling}.

Denoting by $\ueff$ the effective membrane potential (i.e., the average
membrane potential under constant external stimulus other than the synaptic
noise), this yields a sigmoidal activation function $\tilde\sigma(\ueff)$ (see
Fig. \ref{fig:neuralsampling}~D), which can be linearly transformed to the
logistic activation function $\sigma(v)$ to match the abtract model in Sec.\ \ref{sec:neuralsampling}:
\begin{equation}
        \label{lif_to_stoch}
        v = \frac{\ueff-\expect{u}_0}{\alpha} \quad ,
\end{equation}
where $\expect{u}_0$ represents the value of $\ueff$ for which $p(Z=1)=1/2$.
The factor $\alpha$ denotes a scaling factor between the two domains and is equal to $4 \left[ \mathrm{d}\tilde\sigma/\mathrm{d}\ueff(\expect{u}_0) \right]^{-1}$.
From here, a set of parameter translation rules between the abstract and the LIF domain follow, which are explained in more detail in Sec.~\ref{ss:tra}.
Fig.~\ref{fig:neuralsampling}~E shows the result of sampling with LIF neurons from an example Boltzmann distribution together with the target probability values.

\subsection{Characterization of the Auxiliary Neurons}
\label{ss:aux}

In the mathematical model in Section \ref{ss:bn}, the weights between principal and auxiliary RVs are $\Mexc=\infty$ and $\Minh=-\infty$, to ensure a switching of the joint state whenever one of the auxiliary variables changes its assignment. 
In a concrete implementation, infinite weights are unfeasible.
Here, we set the connection strengths $M_{\mathrm{exc},k}=-M_{\mathrm{inh},k}=\gamma \cdot \max\left[\Phi_k\left(\mathbf{z}_k\right)\right]$, where $\gamma$ is a fixed number between 5 and 10.
Neurons with a bias of $M_{\mathrm{exc},k}$ ($M_{\mathrm{inh},k}$) will effectively spike at maximum rate (remain silent), unless driven by afferent neurons with similarly high synaptic weights.

The individual values of the factor $\Phi_k\left(\mathbf{z}_k\right)$ are introduced through the bias of the auxiliary neurons:
\begin{equation}
    \label{impl1_bias}
    a_{X_k^{\mathbf{z}_k}} = \log \left(\mu \frac{\Phi_k\left(\mathbf{z}_k\right)}{\min_{\mathbf{z}_k}\left[\Phi_k\left(\mathbf{z}_k\right)\right]} - 1\right) - L^1(\mathbf{z}_k) \cdot M_{\mathrm{exc},k}
\end{equation}
where the factor $\mu / \min_{\mathbf{z}_k}\left[\Phi_k\left(\mathbf{z}_k\right)\right]$ ensures that the argument of the logarithm stays larger than $0$ for all possible assignments $\mathbf{z}_k$.

Observed variables are clamped to fixed values 0 or 1 by setting the biases of the corresponding principal neurons to very large values ($\pm 20$), to ensure that they spike either at maximum rate or not at all.
This mimics the effect of strong excitatory or inhibitory stimulation.

\subsection{Parameter Translation between Distributions and Networks}
\label{ss:tra}

In the LIF domain, the bias $b$ can be set by changing the leak potential $\El$ such that the neuron is active with $\sigma(b)$ for $\v{Z}_{\backslash k}=\v{0}$:
\begin{equation}
    \label{bias_lif}
    \El = \ueff^{b} \frac{\gl}{\expect\gtot} = \left( \alpha b + \expect{u}_0 \right)\frac{\gl}{\expect\gtot} \quad ,
\end{equation}
where $\gtot$ represents the total synaptic conductance and $\ueff^{b}$ is the effective membrane potential that corresponds to the bias $b$: $\tilde\sigma(\ueff^{b})=\sigma\left(b\right)$.

For the translation of synaptic weights, we use the approximate PSP shape of an LIF neuron with conductance-based synapses in the high-conductance state (HCS)  \citep{petrovici13lifsampling}:
\begin{align}
        \label{condPSP}
        u_\mathrm{PSP}(t) & \approx \frac{w_{ki} \left(E_k^\mathrm{rev} - \expect{u}\right)}{ 
        C_\mathrm{m} \cdot \left(\frac{1}{\tau_\mathrm{syn}} - \frac{1}{\tau_\mathrm{eff}}\right)} \\
        & \quad \left[\exp\left(-\frac{t - t_\mathrm{spike}}{\tau_\mathrm{eff}}\right) - 
        \exp\left(-\frac{t - t_\mathrm{spike}}{\tau_\mathrm{syn}}\right) \right] \quad , \nonumber
\end{align}
where, $w_{ki}$ denotes the synaptic weight from neuron $i$ to neuron $k$ and $\taueff=\Cm/\expect{\gtot}$ the effective membrane time constant.
For both the LIF domain and the abstract domain, a presynaptic spike is intended to have the same impact on the postsynaptic neuron, which is approximately realized by matching the average value of the LIF PSPs within a refractory period with the theoretically optimal constant value:
\begin{equation}
        \label{matching_psps}
        \underbrace{\frac{1}{\alpha} \int_0^{\tau_\mathrm{ref}} u_\mathrm{PSP}(t) \, \mathrm dt}_{\text{LIF neuron}} \overset{!}{=} \underbrace{W_{ki} \cdot \tau_\mathrm{ref}}_{\text{abstract model}} \; .
\end{equation}
Evaluating the integral in Eq.~\ref{matching_psps} yields the weight translation factor between the abstract and the LIF domain: $w_{ki} = \beta \cdot W_{ki}$, where
\begin{align}
        \label{weight_translation_factor}
        \beta & = \frac{\alpha C_\mathrm{m} \tau_\mathrm{ref}\left(\frac{1}{\tau_\mathrm{syn}} - \frac{1}{\tau_\mathrm{eff}}\right)}
        {E_k^\mathrm{rev} - \expect{u}} \\
        & \quad \left[\tau_\mathrm{syn} \left(\mathrm e^{-\frac{\tau_\mathrm{ref}}{\tau_\mathrm{syn}}} - 1\right) - 
        \tau_\mathrm{eff} \left(\mathrm e^{-\frac{\tau_\mathrm{ref}}{\tau_\mathrm{eff}}} - 1\right) \right]^{-1} \quad . \nonumber
\end{align}

Fig.~\ref{fig:kkires}A shows the shape of such an LIF PSP with parameter values taken from Tab.~\ref{tab:cobalif}.
The shape is practically exponential, due to the extremely short effective membrane time constant in the HCS.
We will later compare the performance of the LIF implementation to two implementations of the abstract model from Sec.~\ref{sec:neuralsampling}:
neurons with theoretically optimal rectangular PSPs of duration $\tauref$, the temporal evolution of which is defined as
\begin{equation}
        u(t) = \left\{\begin{array}{ll}
                            1 \quad & \quad \text{if $0 < t < \tauref$} \; ,\\
                            0 \quad & \quad \text{otherwise} \end{array}\right.
\end{equation}
and neurons with alpha-shaped PSPs with the temporal evolution
\begin{equation}
    u(t) = \left\{\begin{array}{ll}
            q_1 \cdot \left[\mathrm{e} \cdot
            \left(\frac{t}{\tau_\alpha}+t_1\right) \cdot
            \exp\left(-\frac{t}{\tau_\alpha} - t_1\right) -
        0.5\right] \quad \\ \quad \text{if $0 < t < (t_2-t_1)\tau_\alpha$} \; ,\\
    0 \quad \text{otherwise} \; . \end{array}\right.
\end{equation}
Here, $t_1$ and $t_2$ are the points in time where the alpha kernel is $\mathrm{e} \cdot t \cdot \exp(-t) = 0.5$.
The value $q_1 = \unit[2.3]{}$ is a scaling factor and $\tau_\alpha = \unit[17]{ms} \cdot \frac{\tauref}{\unit[30]{ms}}$ is the time constant of the kernel \citep{pecevski11}.

In the abstract neural model, by definition, the rectangular PSPs can not superpose, since their width is identical to the refractory period of the neurons.
In LIF neurons, PSPs do not end abruptly, possibly leading to (additive) superpositions, and thereby to deviations from the target distribution.
To counteract this effect, we have used the Tsodyks-Markram model of short-term synaptic plasticity \citep{Tsodyks97}.
Setting the facilitation constant $\tau_\mathrm{facil}=0$ leads to $u_{n+1}=U_0$. 
With the initial utilization parameter $U_0=1$ and the recovery time constant $\tau_\mathrm{rec}=\tausyn$, the parameter $R$, which describes the recovery of the synaptic strength after the arrival of an action potential, yields
\begin{equation}
       \label{workaround}
       R_{n+1} = 1 - \exp\left(-\frac{\Delta t}{\tau_\mathrm{syn}}\right) \, ,
\end{equation} 
where $\Delta t$ is the time interval between the $n^{th}$ and the $(n+1)^{th}$ afferent spike.
The condition in Eq. \ref{workaround} is equivalent to a renewing synaptic conductance, which, due to the fast membrane in the HCS, is in turn equivalent to renewing PSPs. 

\subsection{Performance Improvement via a Superposition of LIF PSP Kernels}
\label{ss:mlif}

\begin{figure*}[tbp]
    \centering
    \includegraphics[width=1.0\linewidth]{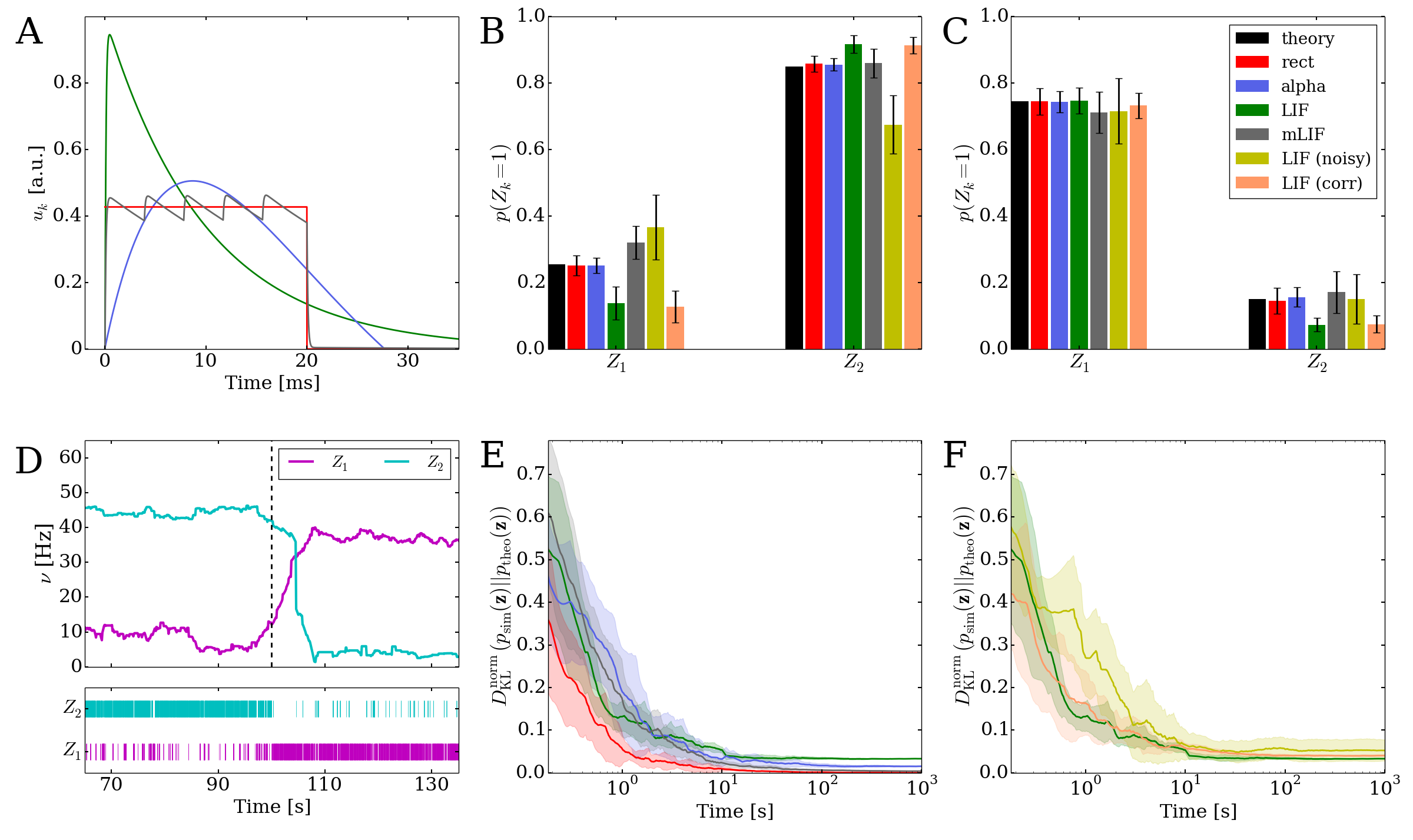}
    \caption{
         Comparison of the different implementations of the Knill-Kersten graphical model (Fig.~\ref{fig:kki}): LIF (green), LIF with noised parameters (yellow), 
         LIF with small cross-correlations between noise channels (orange), mLIF PSPs mediated by a superposition of LIF PSP kernels (gray),
         abstract model with alpha-shaped PSPs (blue), abstract model with rectangular PSPs (red) and analytically calculated (black).
        The error bars for the noised LIF networks represent the standard error over 10 trials with different noised parameters.
        All other error bars represent the standard error over 10 trials with identical parameters.
        (\textbf{A}) Comparison of the four used PSP shapes.
        (\textbf{B, C}) Inferred marginals of the hidden variables $Z_1$ and $Z_2$ conditioned on the observed (clamped) states of $Z_3$ and $Z_4$.
        In B, $(Z_3, Z_4) = (1,1)$.
        In C, $(Z_3, Z_4) = (1,0)$.
        The duration of a single simulations is \unit[10]{s}.
        (\textbf{D}) Marginal probabilities of the hidden variables reacting to a change in the evidence $Z_4 = 1 \to 0$.
        The change in firing rates (top) appears slower than the one in the raster plot (bottom) due to the smearing
        effect of the box filter used to translate spike times into firing rates.
        (\textbf{E, F}) Convergence towards the unconstrained equilibrium distributions compared to the target distribution.
        In D, the performance of the four different PSP shapes from A is shown.
        The abstract model with rectangular PSPs converges to $\DKL=0$, since it is guaranteed to sample from the correct distribution in the limit $t \to \infty$.
        In E, the performance of the three different LIF implementations is shown.
        \label{fig:kkires}
    }
\end{figure*}

The difference in PSP shapes between the LIF domain and the theoretically optimal abstract model is the main reason why the direct translation to LIF networks causes slight deviations from the target probability distribution.
The sometimes strong interaction involved in the expansion of Bayesian networks into Boltzmann machines (see Eq.~\ref{weight_matrix_v}) leads to a large overshoot of the membrane potential at the arrival of a PSP and a nonzero PSP tail beyond $t=t_\mathrm{spike}+\tauref$ (see Fig.~\ref{fig:kkires}~A).

\begin{figure}[tbp]
    \centering
    \includegraphics[width=0.5\linewidth]{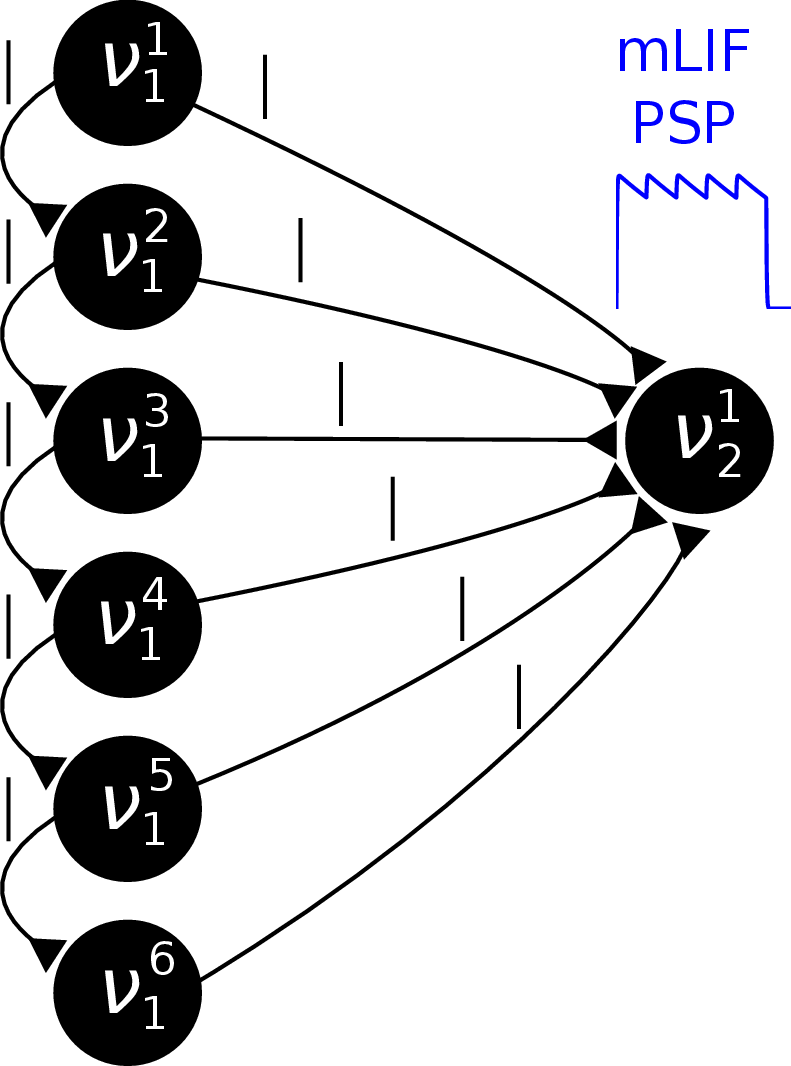}
    \caption{
    In order to establish a coupling which is closer to the ideal one (rectangular PSP),
        the following network structure was set up: Instead of using one principal neuron
        $\nu$ per RV, each RV is represented by a neural chain. 
    In addition to the network connections imposed by the translation of the modeled Bayesian graph, feedforward connections between the neurons in this chain are also generated.
    Furthermore, each of the chain neurons projects onto the first neuron of the postsynaptic interneuron chain 
    (here: all connections from $\nu_1^i$ to $\nu_2^1$).
        By choosing appropriate synaptic efficacies and delays, the chain generates a superposition of 
single PSP kernels that results in a sawtooth-like shape which is closer to the desired rectangular shape than a single PSP.
        \label{fig:improvedPSP}
    }
\end{figure}

In order to reduce this discrepancy, we replaced the single-PSP-interaction between pairs of neurons by a superposition of LIF PSP kernels.
For this, we replaced the single neuron that coded for an RV by a chain of neurons (see Fig.~\ref{fig:improvedPSP}).
In this setup, the first neuron in a chain is considered the ``main'' neuron, and only the spikes it emits are considered to encode the state $z_k=1$.
However, all neurons from a chain project onto the main neuron of the chain representing a related RV.
This neuron then registers a superposition of PSPs, which can be adjusted (e.g., with the parameter values from Tab.~\ref{tab:cobamlif}) to closely approximate the ideal rectangular shape by appropriately setting synaptic weights and delays within as well as between the chains.
In particular, the long tail of the last PSP is cut off by setting the effect of the last neuron in the chain to oppose the effect of all the others (e.g., if the interaction between the RVs is to be positive, all neurons in the chain project with excitatory synapses onto their target, while the last one has an inhibitory outgoing connection).
While this implementation only scales the number of network components (neurons and synapses) linearly with the chosen length of the chains, it improves the sampling results significantly (Fig.~\ref{fig:kkires}~B,~C,~E gray bars/traces).


\section{Results}
\label{s:res}

In the Methods Section, we have provided a comprehensive description of the translation of arbitrary Bayesian graphs to networks of LIF neurons.
Now, we apply these networks to several well-studied cognitive inference problems and study their robustness to various types of substrate imperfections.

\subsection{Bayesian Model of the Knill-Kersten Illusion}
\label{ss:kki}

Fig.~\ref{fig:kki} illustrates the translation of the Bayesian graph describing the well-studied Knill-Kersten illusion \citep{knill91} to the LIF domain.
\begin{newstuff}
Panel A shows the visual stimuli consisting of two geometrical objects, both of which are composed of two identical 3D shapes (two cylinders and two cubes, respectively).
Both stimuli feature the same shading profile in the horizontal direction, but differ in their contours.
The perception of the reflectance of each stimulus is influenced by the perceived 3D shape:
In the case of a flat surface (cubes), the right object appears brighter than the left one.
This perceived change in reflectance does not happen in the case of the cylinders.
A cylindrical shape is therefore said to \textit{explain away} the shading profile, while a cuboid shape does not, therefore leading the observer to the assumption of a jump in reflectance.
\end{newstuff}

We have chosen this experiment since it has been thoroughly studied in literature and it has a rather intuitive Bayesian representation.
More importantly, it features some essential properties of Bayesian inference, such as higher-order dependencies within groups of RVs and the ``explaining away'' effect.
The underlying Bayesian model consists of four RVs: $Z_1$ (reflectance step versus uniform reflectance), $Z_2$ (cylindrical versus cuboid 3D shape), $Z_3$ (sawtooth-shaped versus some other shading profile) and $Z_4$ (round versus flat contour).
The network structure defines the decomposition of the underlying probability distribution:
\begin{equation}
    \label{VPEjoint}
    p(Z_1,Z_2,Z_3,Z_4) = p(Z_1)\, p(Z_2)\, p(Z_3 | Z_1, Z_2)\, p(Z_4 | Z_2) \quad .
\end{equation}
The inference problem consists in estimating the relative reflectance of the objects given the (observed) contour and shading.
Analytically, this would require calculating $p(Z_1 | Z_3=1, Z_4=0)$ for the cuboid shapes and $p(Z_1 | Z_3=1, Z_4=1)$ for the cylindrical ones.

Fig.~\ref{fig:kkires} shows the behavior of the LIF network that represents this inference problem.
When no variables are clamped, the network samples freely from the unconstrained joint distribution over the four RVs.
The performance of the network, i.e., its ability to sample from the target distribution, is quantified by the Kullback-Leibler (KL) divergence between the target and the sampled distribution
normalized by the entropy of the target distribution:
\begin{equation}
    \DKL(q||p) = \frac{D_\mathrm{KL}(q||p)}{H(p)} \; ,
\end{equation}
with the KL divergence between the sampled distribution $q$ and the target distribution $p$
\begin{equation}
    D_\mathrm{KL}(q||p) = \sum_\mathbf{z} q(\mathbf{z}) \log\left(\frac{q(\mathbf{z})}{p(\mathbf{z})}\right)
\end{equation}
and the entropy of the target distribution $p$
\begin{equation}
    H(p) = - \sum_\mathbf{z} p(\mathbf{z})
    \log\left[p(\mathbf{z})\right]\; .
\end{equation}
When presented with the above inference problem the LIF network performs well at sampling from the conditional distributions $p(Z_1 | Z_3, Z_4)$ 
(Fig. \ref{fig:kkires}~B,~C). When the stimulus is changed during the simulation, the optical illusion,
i.e., the change in the inferred (perceived) 3D shape and reflectance profile,
is clearly represented by a change in firing rates of the corresponding
principal neurons (Fig. \ref{fig:kkires}~D).
For each point in time, the rate is determined by convolution of the spike train with a rectangular kernel
\begin{equation}
    \kappa(t) = \left\{\begin{array}{ll}
    \unit[1/8]{Hz} \quad & \quad \text{for $\unit[-8]{s} < t < 0$} \; ,\\
    0 \quad & \quad \text{otherwise} \; . \end{array}\right.
\end{equation}
At $t=\unit[100]{s}$ ({\it red} line), the evidence is switched: $Z_4=1
\rightarrow 0$.
\begin{newstuff}
The network reacts appropriately on the time scale of several seconds, as can be seen in the spike raster plot.
\end{newstuff}

When not constrained by prior evidence, i.e., when sampling from the joint distribution over all RVs, the LIF network settles on an equilibrium distribution that lies close to the
target distribution (Fig.~\ref{fig:kkires}~E,~F, green traces).
For this particular network, the convergence time is of the order of several tens of seconds.

\subsection{Robustness to Parameter Distortions}
\label{ss:pardist}

We further investigated the robustness of our proposed implementation of Bayesian inference with LIF neurons to low levels of parameter noise
(see Tab.~\ref{tab:cobalif}, noisy).
Here, we focus on fixed-pattern noise, which is inherent to the production process of semiconductor integrated circuits and is particularly relevant for analog neuromorphic hardware \citep{mitra2009real,petrovici2014characterization}.
However, such robustness would naturally also benefit in-vivo computation.

Some of the noise (the one affecting the neuron parameters that are not changed when setting weights and biases) can be completely absorbed into the translation rules from Sec.~\ref{ss:lif}.
Once the neurons are configured, their activation curves can simply be measured, allowing a correct transformation from the abstract to the LIF domain.
However, while the neurons remain the same between different simulation runs, the weights and biases may change depending on the implemented inference problem and are still subject to noise.
Nevertheless, even with a noise level of 10\% on the weights and biases, the LIF network still produces useful predictions (Fig.~\ref{fig:kkires}~B,~C,~F yellow bars/traces).

\subsection{Robustness to Noise correlations}
\label{ss:cross}

The investigated implementation of Bayesian networks ideally requires each neuron to receive independent noise as a Poisson spike train.
When aiming for a hardware implementation of large Bayesian networks, this requirement may become prohibitive due to the bandwidth limitations of any physical back-end.
We therefore examined the the robustness of our LIF networks to small cross-correlations between the Poissonian noise channels of individual neurons.

For both the excitatory and the inhibitory background pools, we induced pairwise noise correlations by allowing neurons within the network to share 10\% of their background Poisson sources. 
The controlled cross-correlation of 10\% between noise channels is achieved in the following way:
each neuron receives Poisson background from three shared and seven private Poisson spike trains.
The excitatory and inhibitory noise of each individual neuron remained uncorrelated in order to leave its activation function (Eq.~\ref{lif_activation_function}) unaltered.
Each of shared sources projects onto exactly two neurons in order to prevent higher-order correlations.
The single Poissonian spike trains have a firing rate of $\nu/10$, such that their superposition is also Poisson, with the target firing rate of $\nu$.
With this setup, we were able to verify that small pairwise correlations in the background noise do not significantly reduce the ability of the LIF network to produce useful predictions (Fig.~\ref{fig:kkires}~B,~C,~F orange bars/traces).

\begin{figure*}[tbp]
    \centering
    \includegraphics[width=1.0\linewidth]{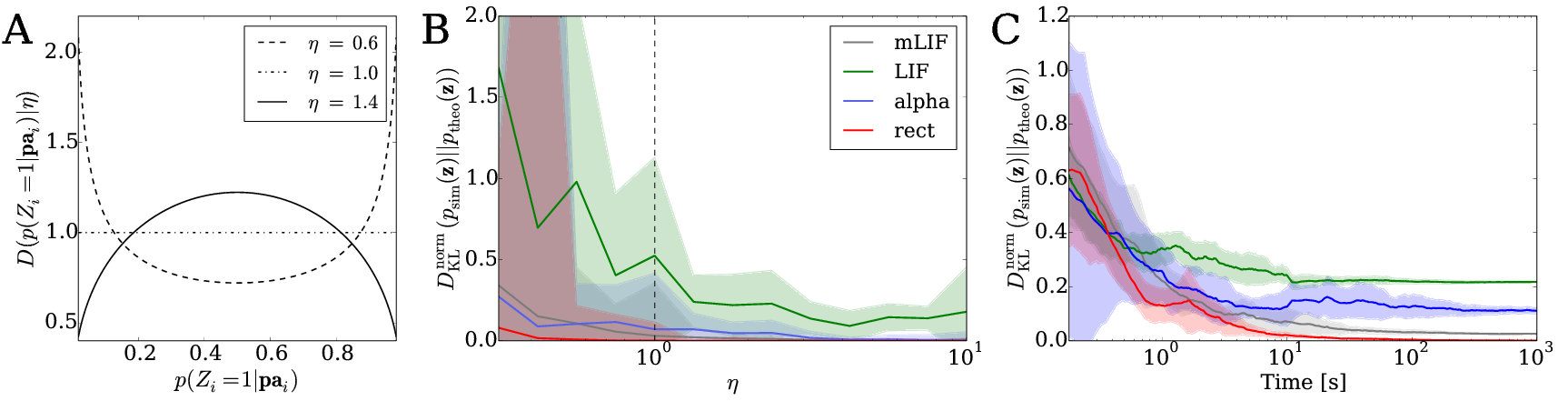}
    \caption{
     Sampling from random distributions over 5 RVs with different networks: LIF (green), mLIF (gray), 
     abstract model with alpha-shaped PSPs (blue) and abstract model with rectangular PSPs (red).
    (\textbf{A}) Distributions for different values of $\eta$ from which conditionals are drawn.
    (\textbf{B}) $\DKL$ between the equilibrium and target distributions as a function of $\eta$.
        The error bars denote the standard error over 30 different random graphs drawn from the same distribution.
    (\textbf{C}) Evolution of the $\DKL$ over time for a sample network drawn from
    the distribution with $\eta=1$. Error bars denote the standard error over 10 trials.
        \label{fig:dklalpha}
    }
\end{figure*}
\subsection{General Bayesian Networks}
\label{ss:gen}

In order to study the general applicability of the proposed approach, we quantified the convergence behavior of LIF networks generated from random Bayesian graphs.
Here, we used a method proposed in \citet{Ide02randomgeneration} to generate random Bayesian networks with $K$ binary RVs and random conditional probabilities.
The algorithm starts with a chain graph $Z_1 \to Z_2 \to \dots \to Z_K$ and runs for $N$ iterations.
In each iteration step, random RV pairs $(Z_i,Z_j)$ with $i>j$ are created.
If the connection $Z_i \to Z_j$ does not exist, it is added to the graph, otherwise it removed, with two constraints:
any pair of nodes may not have more than 7 connections to other nodes and the procedure may not disconnect the graph.
For every possible assignment of $\pa_i$, the conditional probabilities $p_i^{\pa_i}:=p(Z_i=1 | \pa_i)$ are drawn from a second-order Dirichlet distribution
\begin{equation}
    \label{dirichletdistr}
    D(p_i^{\pa_i}, \eta_1, \eta_2) = \frac{1}{B(\eta_1, \eta_2)} (p_i^{\pa_i})^{\eta_1-1} (1-p_i^{\pa_i})^{\eta_2-1} \quad ,
\end{equation}
with the multinomial Beta function
\begin{equation}
    \label{betafunction}
    B(\eta_1, \eta_2) = \frac{\prod^2_{i=1} \Gamma\left(\eta_i\right)}{\Gamma\left(\sum^2_{i=1}\eta_i\right)} \quad ,
\end{equation}
where $\Gamma(\cdot)$ denotes the gamma function.
We chose the parameters $\eta_1=\eta_2=:\eta$ in order to obtain a symmetrical distribution. 
Fig.~\ref{fig:dklalpha}~A shows three examples of a symmetrical two-dimensional Dirichlet distribution.
A larger $\eta$ favors conditional probabilities which are closer to $0.5$ than to the boundaries $0$ and $1$. 

We implemented Bayesian networks with $K=5$ RVs running for $N=50000$ iterations.
The random graphs were then translated to sampling neural networks, both with abstract model neurons and LIF neurons.
The performance was tested for sampling from the unconstrained joint distributions over the $5$ RVs. 
In the simulations, we varied $\eta$ between $0.3$ and $10$ and created $30$ random Bayesian graphs for each $\eta$.
Each network was then run for a total duration of $\unit[100]{s}$.

Fig.~\ref{fig:dklalpha}~B illustrates the average sampling results for the
different PSP shapes as a function of the ''extremeness'' of the randomized
conditional probabilities, which is reflected by the parameter $\eta$.
For larger $\eta$, conditionals cluster around 0.5 and the RVs become more independent, making the sampling task easier and therefore improving the sampling performance.
The curves show the median of the $\DKL$ between sampled and target distributions of the $30$ random Bayesian graphs.
The shaded regions denote the standard error.
Overall, the LIF networks perform well, capturing the main modes of the target distributions.

Fig.~\ref{fig:dklalpha}~C shows the temporal evolution of the $\DKL$ between sampled and target distributions for a sample
Bayesian network drawn from the distribution with $\eta=\unit[1]{}$ that lied close to the $\DKL$ median in Fig.~\ref{fig:dklalpha}~B.
The curves illustrate the average results of $10$ simulations, while the shaded regions denote the standard error.

As with the Bayesian model of the Knill-Kersten illusion, the main cause of the remaining discrepancy is the difference in PSP shapes between the LIF domain and the theoretically optimal abstract model.
A modification of the RV coupling by means of the neuron chains described in Sec.~\ref{ss:mlif} leads to a significant improvement of the sampling results for arbitrary Bayesian networks (Fig.~\ref{fig:dklalpha}~B,~C gray traces).

\section{Discussion}

\label{s:con}

In this article, we have presented a complete theoretical framework that allows the translation of arbitrary probability distributions over binary RVs to networks of LIF neurons.
We build upon the theory from \citet{buesing2011neural} and \citet{pecevski11} and extend their work to a mechanistic neuron model widely used in computational neuroscience based on the approach in \citet{petrovici13lifsampling}.
In particular, we make use of the conductance-based nature of membrane dynamics to enable fast reponses of neurons to afferent stimuli.

We have demonstrated how networks of conductance-based LIF neurons can represent probability distributions in arbitrary spaces over binary RVs and can perform stochastic inference therein.
By comparing our proposed implementation to the theoretically optimal, abstract model we have shown that the LIF networks produce useful results for the considered inference problems.
Our framework allows a comparatively sparse implementation in neural networks, both in terms of the absolute number of neurons as well as considering energy expenditure for communication, since state switches are encoded by single spikes.
This compares favorably with other implementations of inference with LIF neurons, based on e.g. firing rates or reservoir computing \citep{SteimerMD09}.

The main cause for the deviations of the LIF equilibrium distributions from the target distributions lie in the shape of synaptic PSPs.
We shave shown how a more complex coupling mechanism based on interneuron chains can improve inference by allowing a more accurate translation of target distributions to networks of LIF neurons.
This kind of interaction provides a connection to other well-studied models of chain-based signal propagation in cortex \citep{diesmann1999stable,kremkow2010functional,petrovici2014characterization}.
A similar interaction kernel shape can be achieved by multiple synapses between two neurons (multapses) lying at different points along a dendrite, causing their PSPs to arrive at the soma with different delays.
From a computational point of view, the increased sampling performance due to this coupling mechanism only comes at the cost of linearly increasing network resources.

\begin{newstuff}
    An explicit goal of our theoretical framework was to establish a rigorous translation of abstract models of Bayesian inference to neural networks based on mechanistic neuron models that are commonplace in computational neuroscience.
    Our particular formulation uses LIF neurons, but a translation to similar integrate-and-fire spiking models such as AdEx \citep{brette2005adaptive} is straightforward.
    An equivalent formulation for more biological models such as Hodgkin-Huxley \citep{hodgkin52quantitative} is feasible in principle, but non-trivial, mostly due to the fact that the Hodgkin-Huxley model inherently incorporates a form of relative refractoriness.
    A study of neural sampling with relative refractoriness does exist \citep{buesing2011neural}, but how the abstract model is mappable to Hodgkin-Huxley dynamics remains an open question.
    
    Another important issue concerns how the structure of these networks can be learned from data samples through synaptic plasticity mechanisms such as STDP.
    The auxiliary subnetworks required by our model are conceptually equivalent to the well-studied winner-take-all (WTA) motif.
    The emergence of such structures for discriminative tasks based on both supervised and unsupervised STDP protocols has been the subject of recent investigations \citep{nessler13,kappel2014stdp}.
    The application of these protocols to networks of integrate-and-fire neurons in general and to our implementation of Bayesian networks in particular is the subject of ongoing research.

    Concerning the practical application of our model, we have studied its robustness to small levels of parameter noise as well as weak correlations between the noise channels of individual neurons.
    We were able to show that imperfections of the physical substrate of the neuronal implementation, be it biological or artificial, can be well tolerated by our networks.
    Further improvement of the robustness towards parameter noise, as well as a higher degree of biological plausibility, can be achieved by implementing individual RVs as populations of neurons, as has been recently proposed by \citet{legenstein14}.
\end{newstuff}
We therefore regard our model as a promising candidate for implementation on neuromorphic devices, which can augment these already efficient networks by providing a fast, low-power emulation substrate.
Implemented on such substrates, our networks can serve as a basis for machine learning algorithms, thereby facilitating the development of, e.g., autonomous robotic learning agents.

\bibliography{refs}

\section*{Appendix}

Table \ref{tab:cobalif} lists the standard neuron and network parameters used in this paper.
Parameter values for the experiments with mLIF PSPs are shown in Table \ref{tab:cobamlif}.

\begin{table*}
\centering
\begin{tabular}{lrr}\toprule
	{\bf LIF parameter} & {\bf standard} & {\bf noisy}\\
	\colrule
        Resting membrane potential $V_\textnormal{rest}$ & $\overline{V_k^b}$ & $\overline{V_k^b} \pm \unit[2.0]{mV}$\\
        Capacity of the membrane $C_\textnormal{m}$ & $\unit[0.2]{nF}$ & $\unit[0.2]{nF}$\\
        Membrane time constant $\tau_\textnormal{m}$ & $\unit[0.1]{ms}$ & $\unit[(1.0 \pm 0.1)]{ms}$\\
        Duration of refractory period $\tau_\textnormal{ref}$ &
        $\unit[20.0]{ms}$ & $\unit[(20.0 \pm 1.0)]{ms}$\\
        Excitatory synaptic time constant
        $\tau^\textnormal{syn,exc}$ & $\unit[10.0]{ms}$ & $\unit[(20.0 \pm 2.0)]{ms}$\\
        Inhibitory synaptic time constant
        $\tau^\textnormal{syn,inh}$ & $\unit[10.0]{ms}$ & $\unit[(20.0 \pm 2.0)]{ms}$\\
        Reversal potential for excitatory input
        $E^\textnormal{rev,exc}$ & $\unit[0.0]{mV}$ & $\unit[(0.0 \pm 2.0)]{mV}$\\
        Reversal potential for inhibitory input
        $E^\textnormal{rev,inh}$ & $\unit[-100.0]{mV}$ & $\unit[(-100.0 \pm 2.0)]{mV}$\\
        Spike threshold $V_\textnormal{th}$ & $\unit[-50.0]{mV}$ & $\unit[(-50.0 \pm 0.5)]{mV}$\\
        Reset potential after a spike $V_\textnormal{reset}$ &
        $\unit[-53.0]{mV}$ & $\unit[(-53.0 \pm 0.5)]{mV}$\\
        Utilization of synaptic efficacy $U_0$ & $\unit[1.0]{}$ & $\unit[1.0]{}$\\
        Recovery time constant $\tau_\textnormal{rec}$ & $0.99 \cdot \tausyn$ & $0.99 \cdot \tausyn$\\
        Facilitation time constant $\tau_\textnormal{facil}$ & $\unit[0.0]{ms}$ & $\unit[0.0]{ms}$\\
        Excitatory/inhibitory Poisson input rate $\nusyn$ &
        $\unit[400.0]{Hz}$ & $\unit[5000.0]{Hz}$\\
        Excitatory/inhibitory background weight $\wsyn$ &
        $\unit[0.002]{\mu S}$ & $\unit[0.001]{\mu S}$ \\
        Synaptic delays & $\unit[0.1]{ms}$ & $\unit[1.2]{ms}$\\
    \colrule
        {\bf Boltzmann machines: Parameter} & {\bf standard} & {\bf noisy}\\
	\colrule
        $W_{ij}$ & $W_{ij}$ & $\epsilon \cdot W_{ij}$ \\
        $b_i$ & $b_i$ & $\epsilon \cdot b_i$ \\
 	$\gamma$ (Equation 16) & $10$ & $5$\\
	$\mu$ (Equation 17) & $1+10^{-4}$ & $1+10^{-4}$\\
\botrule
\end{tabular}
\caption{\label{tab:cobalif} Standard neuron and network parameters used in this paper. The network parameter $\epsilon$ denotes a sample from the uniform distribution $\unif(0.9,1.1)$.}
\end{table*}

\begin{table*}
\centering
\begin{tabular}{lr}\toprule
	\textbf{Parameters of the first chain neuron} & \\
	\colrule
	Capacity of the membrane $C_\textnormal{m}$ & \unit[0.2]{nF} \\
        Membrane time constant $\tau_\textnormal{m}$ & \unit[0.1]{ms} \\
        Duration of refractory period $\tau_\textnormal{ref}$ & \unit[29.5]{ms} \\
        Decay time of the excitatory synaptic conductance $\tau_\textnormal{syn,exc}$ & \unit[30.0]{ms} \\
        Decay time of the inhibitory synaptic conductance $\tau_\textnormal{syn,inh}$ & \unit[30.0]{ms} \\
        Reversal potential for excitatory input $E_\textnormal{exc}^\textnormal{rev}$ & \unit[0.0]{mV} \\
        Reversal potential for inhibitory input $E_\textnormal{inh}^\textnormal{rev}$ & \unit[-100.0]{mV} \\
        Spike threshold $V_\textnormal{th}$ & \unit[-50.0]{mV} \\
        Reset potential after a spike $V_\textnormal{reset}$ & \unit[-50.01]{mV} \\
	\colrule
	\textbf{Parameters of the remaining chain neurons} & \\
	\colrule
	Capacity of the membrane $C_\textnormal{m}$ & \unit[0.2]{nF} \\
        Membrane time constant $\tau_\textnormal{m}$ & \unit[0.1]{ms} \\
        Duration of refractory period $\tau_\textnormal{ref}$ & \unit[29.3]{ms} \\
        Decay time of the excitatory synaptic conductance $\tau_\textnormal{syn,exc}$ & \unit[2.0]{ms} \\
        Decay time of the inhibitory synaptic conductance $\tau_\textnormal{syn,inh}$ & \unit[2.0]{ms} \\
        Reversal potential for excitatory input $E_\textnormal{exc}^\textnormal{rev}$ & \unit[0.0]{mV} \\
        Reversal potential for inhibitory input $E_\textnormal{inh}^\textnormal{rev}$ & \unit[-100.0]{mV} \\
        Spike threshold $V_\textnormal{th}$ & \unit[-50.0]{mV} \\
        Reset potential after a spike $V_\textnormal{reset}$ & \unit[-52.3]{mV} \\
        Resting membrane potential $V_\textnormal{rest}$ & \unit[-52.3]{mV} \\
	\colrule
	\textbf{Parameters of the neural chain} & \\
	\colrule
        Number of chain neurons & 6 \\
	Delay: sampling $\rightarrow$ sampling neuron & \unit[0.1]{ms} \\
	Delay: sampling $\rightarrow$ forwarding neuron & \unit[5.8]{ms} \\
	Delay: forwarding $\rightarrow$ sampling neuron & \unit[0.1]{ms} \\
	Delay: forwarding $\rightarrow$ forwarding neuron & \unit[5.8]{ms} \\
	Delay: forwarding $\rightarrow$ last forwarding neuron & \unit[5.9]{ms} \\
	Weight: sampling $\rightarrow$ sampling neuron & $w$ \\
	Weight: sampling $\rightarrow$ forwarding neuron & \unit[0.16]{$\mu$S} \\
	Weight: forwarding $\rightarrow$ sampling neuron & $0.180 \cdot w$ \\
	Weight: last forwarding $\rightarrow$ sampling neuron & $-0.815 \cdot w$ \\
        Weight: forwarding $\rightarrow$ forwarding neuron & \unit[0.16]{$\mu$S} \\
\botrule
\end{tabular}
\caption{\label{tab:cobamlif} Parameters of the interneuron chain of 6 neurons, which are used to generate the mLIF PSP.}
\end{table*}

\end{document}